\documentclass[11pt]{article}

\usepackage[utf8]{inputenc}
\usepackage[T1]{fontenc}
\usepackage{amsmath,amssymb}
\usepackage{graphicx}
\usepackage{booktabs}
\usepackage[margin=1in]{geometry}
\usepackage{hyperref}
\usepackage{natbib}
\usepackage{xcolor}

\title{Estimating Individualized Treatment Effects in Acute Ischemic Stroke
with Causal Transformation Models (TRAM-DAG):\\
A Multi-Centre Observational Study with External RCT Validation}

\author{
Lisa Herzog$^{3}$,\;
Oliver D\"urr$^{1,2}$,\;
Pascal B\"uhler$^{1}$,\;\\
Hakim Baazaoui$^{3}$,\;
Julian Dese\"o$^{3}$,\;
Susanne Wegener$^{3,4}$,\;
Beate Sick$^{1,4}$,\;
\\[0.5em]
{\small $^1$\,Thurgau Institute for Digital Transformation (TIDIT), Kreuzlingen, Switzerland}\\
{\small $^2$\,HTWG Konstanz -- University of Applied Sciences Konstanz, Germany}\\
{\small $^3$\,Department of Neurology, University Hospital Zurich (USZ), Switzerland}\\
{\small $^4$\,University of Zurich (UZH), Switzerland}
}

\date{\today}

\begin{document}
\maketitle
\begin{abstract}
Personalized medicine in acute ischemic stroke requires moving beyond average
treatment effects (ATE) to individualized treatment effects (ITE) estimates 
to support treatment decisions. In acute 
ischemic stroke, mechanical thrombectomy has been shown to be more effective on 
average than lysis in randomized controlled trials (RCTs), such as the MR CLEAN study.  
We aim to identify which individual patients benefit most from
 mechanical thrombectomy compared to lysis. The outcome of interest is the modified Rankin Scale 
(mRS) at three months, an ordinal measure of functional disability 
(0: no symptoms, 6: death).
We demonstrate that causal transformation models on directed acyclic graphs 
(TRAM-DAG) can be used for ITE estimation after being fitted on observational MAGIC multi-center stroke patient data. To ensure comparability with the MR CLEAN population, which we plan to use for validation, we train the TRAM-DAG on a MAGIC sub-population with NIHSS at admission $\geq6$, corresponding to one inclusion criterion of MR CLEAN.  
The fitted model is then used to estimate individualized treatment effects (ITE) for 
stroke patients in the MR CLEAN population. 
While these ITE estimations cannot be directly confirmed experimentally, we confirmed that their average is  consistent 
with the trial's reported ATE. 
Furthermore, based on the ITE estimates, the trial patients can be correctly ranked   to the 
observed frequency of a good outcome ($mRS\_3m \leq 2$). 
These findings support the use of causal models like TRAM-DAG for 
personalized decision-making in stroke care and highlight their ability to 
bridge the gap between observational evidence and clinical trials.

\end{abstract}

\section{Introduction}
Acute ischemic stroke is a leading cause of disability worldwide. While
mechanical thrombectomy (Intra-Arterial Treatment, IAT; $T=1$) has been
proven effective on average in randomized controlled trials (RCTs), clinical
outcomes vary widely among individuals \citep{mrclean}. Precision medicine aims to provide
each patient with the treatment that maximizes their specific probability
of a good outcome.

However, RCTs are often limited by strict inclusion criteria and may not
reflect the heterogeneous populations seen in daily clinical practice.
Observational data, while more representative, are subject to confounding.
Causal modeling, specifically using Directed Acyclic Graphs (DAGs), allows
us to encode domain knowledge and estimate individualized treatment effects (ITE) from
observational data. In this work, we apply TRAM-DAG \citep{tramdag}, a
flexible and interpretable causal model based on transformation models, to
estimate ITEs in stroke. We validate our approach by comparing its
estimated interventional predictions against real-world RCT data from the MR CLEAN
trial \citep{mrclean}.

\section{Data}
\label{sec:data}
We utilize two distinct datasets: 

1)  MR CLEAN RCT data \citep{mrclean} which serves as external validation data
In MR CLEAN $500$ patients are randomized  either to
mechanical thrombectomy plus usual care ($T=1$, $n=233$) or usual care alone
($T=0$, $n=267$). As shown in Table~\ref{tab:cohorts}, the RCT design ensures
that baseline characteristics like age and NIHSSa are well-balanced across
treatment arms.

2) MAGIC, an observational multi-center stroke patient 
cohort \citep{magic}, which we use as the training data for the TRAM-DAG. 
The  observational MAGIC dataset contains stroke patients from 
five European centers. To bridge the gap between observational MAGIC data and the MR CLEAN RCT data, we restrict ourselves to a MAGIC sub-population with NIHSS at admission $\geq6$ corresponding to one inclusion criterion of MR CLEAN (see Table~\ref{tab:data}).

\begin{table}[h!]
\centering
\caption{Observational MAGIC cohort by centre, before (all patients) and after
the analysis restriction to NIHSSa $\ge 6$.}
\label{tab:data}
\begin{tabular}{lcc}
\hline
Centre & All patients & NIHSSa $\ge 6$ \\
\hline
USB & 584 & 406 \\
USZ & 494 & 324 \\
KSSG & 375 & 249 \\
Egas\_Moniz & 190 & 178 \\
CHUV & 160 & 118 \\
\textbf{Total} & \textbf{1803} & \textbf{1275} \\
\hline
\end{tabular}

\end{table}

Table~\ref{tab:cohorts} summarizes the distribution of variables in 
both cohorts, stratified by treatment.

\begin{table}[h!]
\centering
\caption{Cohort characteristics of the observational MAGIC cohort before
filtering (all) and after restriction to NIHSSa $\ge 6$, compared with the
MR CLEAN RCT. The filter raises MAGIC severity and lowers its good-outcome rate,
moving it toward the RCT. $T=1$ indicates mechanical thrombectomy; $T=0$ is
usual care.}
\label{tab:cohorts}
\resizebox{\textwidth}{!}{\footnotesize
\begin{tabular}{lcccccc}
\toprule
& \multicolumn{2}{c}{MAGIC (all)} & \multicolumn{2}{c}{MAGIC (NIHSSa$\ge6$)} & \multicolumn{2}{c}{MR CLEAN (RCT)} \\
\cmidrule(lr){2-3}\cmidrule(lr){4-5}\cmidrule(lr){6-7}
Variable & $T=0$ ($N=610$) & $T=1$ ($N=1193$) & $T=0$ ($N=280$) & $T=1$ ($N=995$) & $T=0$ ($N=267$) & $T=1$ ($N=233$) \\
\midrule
Age, mean (SD) & 72.4 (13.9) & 71.3 (13.9) & 76.3 (13.2) & 71.9 (13.8) & 65.2 (13.7) & 64.7 (13.9) \\
NIHSSa, median [IQR] & 5 [2--10] & 14 [8--18] & 10 [8--16] & 15 [11--19] & 18 [14--22] & 17 [14--21] \\
mRS\_pre, median [IQR] & 0 [0--1] & 0 [0--1] & 1 [0--2] & 0 [0--1] & 0 [0--0] & 0 [0--0] \\
mRS\_3m, median [IQR] & 2 [1--4] & 2 [1--4] & 3 [2--5] & 3 [1--4] & 4 [3--5] & 3 [2--5] \\
Good outcome (mRS $\le2$) & 60.7\% & 52.6\% & 39.6\% & 47.6\% & 19.1\% & 32.6\% \\
\bottomrule
\end{tabular}
}
\end{table}

\begin{figure}[h!]
\centering
\includegraphics[width=0.95\linewidth]{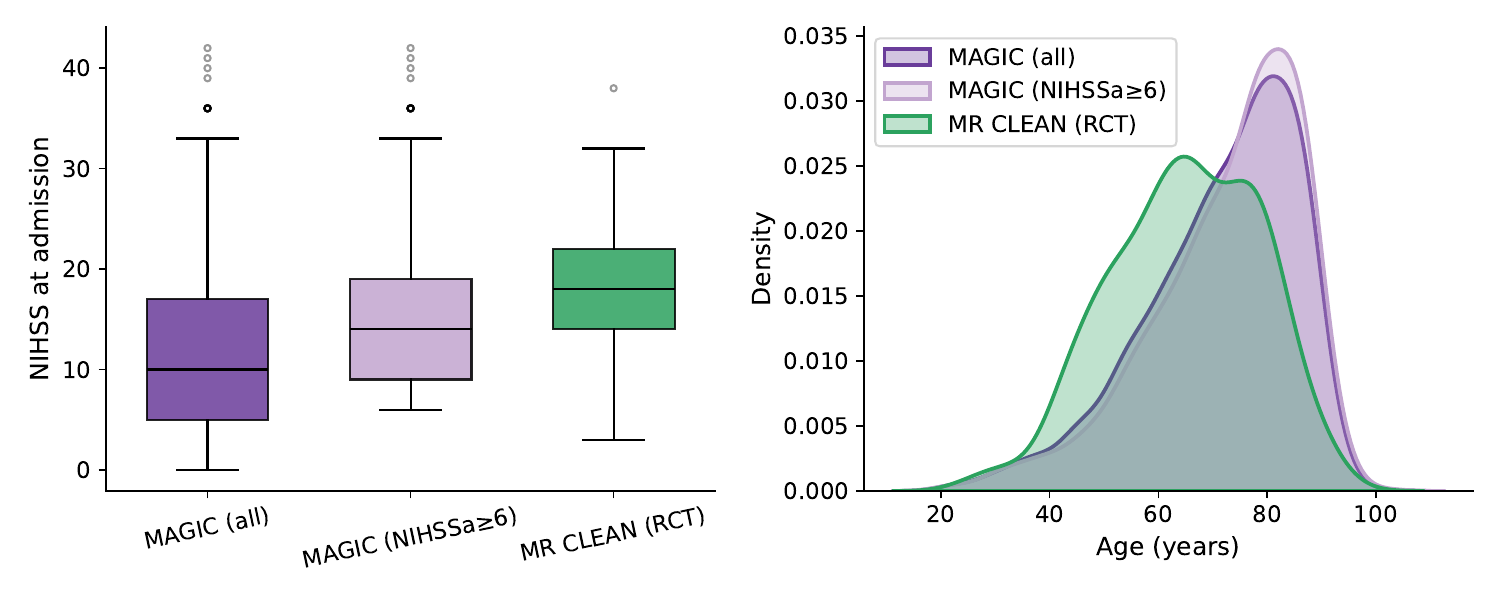}
\caption{Distributions of stroke severity (NIHSS at admission, left) and age
(right) for the observational MAGIC cohort before filtering (purple) and after
restriction to NIHSSa $\ge 6$ (light purple), compared with the MR CLEAN RCT
(green). The NIHSSa $\ge 6$ restriction shifts the MAGIC severity distribution
toward the more severe RCT population (median NIHSS $10\rightarrow14$ versus $18$
in the RCT), while age remains higher in MAGIC.
}
\label{fig:eda}
\end{figure}

The MAGIC and MR CLEAN cohorts differ substantially in their baseline characteristics. Figure~\ref{fig:eda}
compares the distributions of age and stroke severity (NIHSSa). Patients in the
observational MAGIC cohort are generally older but
presented with significantly less severe strokes (lower NIHSSa) compared
to those in the MR CLEAN RCT. After filtering the MAGIC data according to the criteria NIHSSa $\ge 6$, the distribution shift is less strong for NIHSSa but did not decrease for age.  We used this restricted MAGIC data for all results shown in this report.

According to Table~\ref{tab:data}, patients in the MAGIC cohort receiving 
mechanical thrombectomy
($T=1$) had significantly higher stroke severity (median NIHSSa $14$) than
those who did not (median $5$; see Table~\ref{tab:cohorts}) indicating a potential confounding by indication. 
The overall rate of a good functional outcome was much higher in the 
observational MAGIC cohort than in the RCT cohort, which might be due to the 
much higher stroke severity (NIHSSa)
in the RCT cohort compared to the observational cohort (see Figure~\ref{fig:eda}).

\section{Methods}

\subsection{Causal model: TRAM-DAG}
\begin{figure}[h!]
\centering
\includegraphics[width=0.7\linewidth]{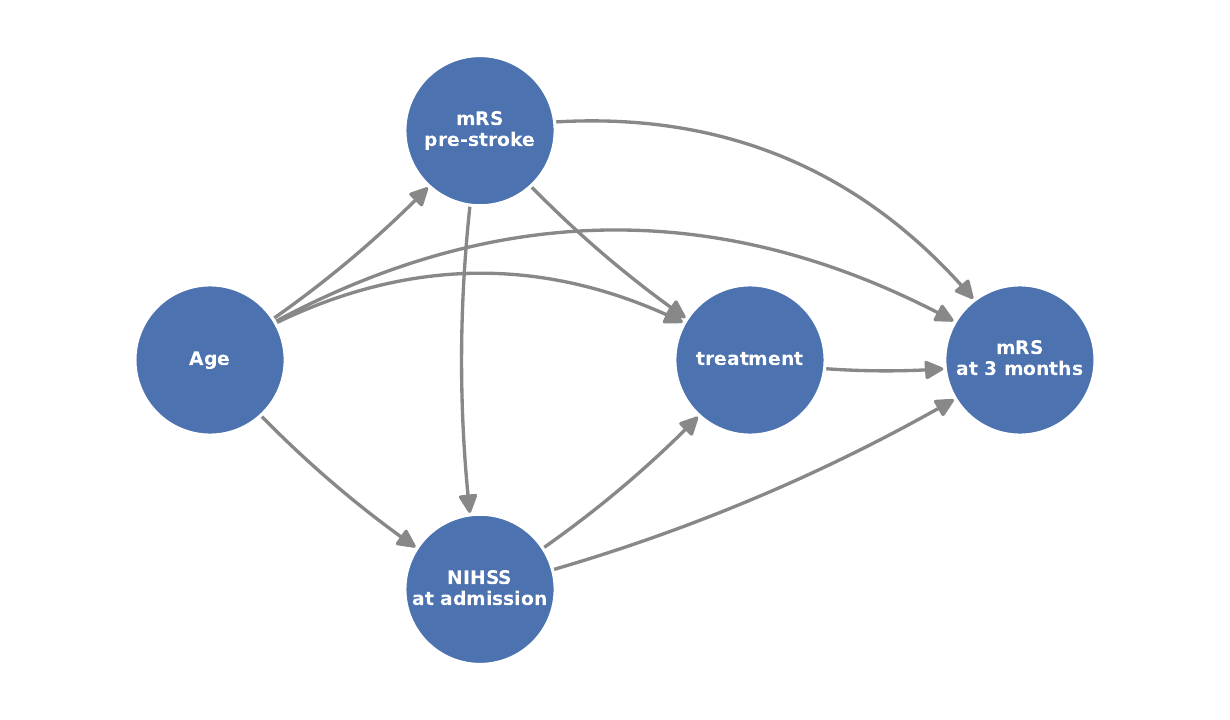}
\caption{This directed acyclic graph (DAG) shows the assumed causal relationships
between the relevant variables to describe the data generation process for stroke patients.
Each arrow indicates a potential direct causal influence. }
\label{fig:dag}
\end{figure}

We represent our model for the causal relationships of the patient variables as a 
DAG (Figure~\ref{fig:dag}). The DAG is based on field knowledge. One causal relation under debate was from mRS pre stroke to NIHSSa. However, we could confirm with a general conditional independence test \citep{glip} omitting this causal relation would not be consistent with the observed data dependencies in the MAGIC data.

\begin{figure}[h!]
\centering
\includegraphics[width=0.6\linewidth]{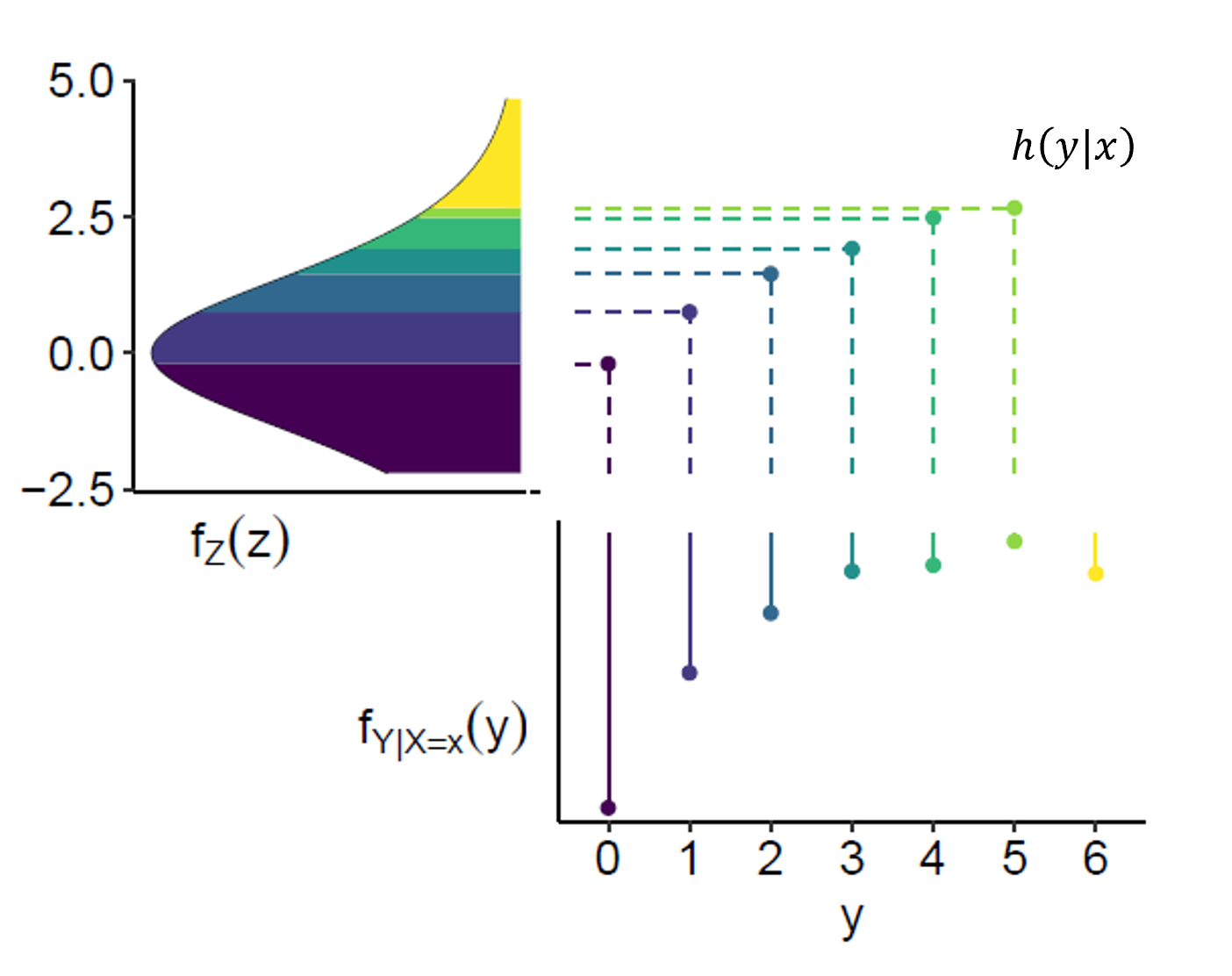}
\caption{TRAM for the outcome y (the ordinal outcome mRS after 3 months) depending
on the value of x (treatment, age, mRS\_pre, NIHSSa). 
The lower right panel shows the predicted probability distribution of the outcome. 
The core is the discrete transformation function
$h(y|x)$ that depends on the values x that are the parent nodes of y.
$h$ cuts the continuous latent logistic distribution (continuous density plot on the left) into discrete intervals.
The probability for 
each mRS class (y) corresponds to the area under the density between two cut points 
on the latent scale.}
\label{fig:cutpoints}
\end{figure}

We fit one transformation model (TRAM) per node \citep{tramdag,ontram}, conditioning
each node on its causal parents. The core of a TRAM is a monotone transformation $h$
that depends on the parent values. For the ordinal outcome mRS\_3m (seven classes) the transformation reduces to a
set of six ordered cut points (see Figure~\ref{fig:cutpoints}). By the structure of $h$ we can tune the interpretability and flexibility of the model. Based on our experience with stroke data, we allowed age and NIHSSa to have a non-linear effect on the logit scale, while all other effects are modeled to be linear.
For details, see \citep{tramdag,ontram}.

\subsection{Training and interventional predictions}

The model is trained on the restricted MAGIC data with NIHSSa $\ge 6$ 
($N=1{,}275$) defined in Section~\ref{sec:data}.

The restricted MAGIC data were split into training and test sets (9:1), and during training, the training negative log-likelihood (NLL) was used as the loss function, with the validation NLL serving as the criterion for early stopping. 
We checked the model fit by sampling from the model and confirmed that the distribution of the
samples and the observed data are consistent (data not shown).

\begin{figure}[h!]
\centering
\includegraphics[width=0.7\linewidth]{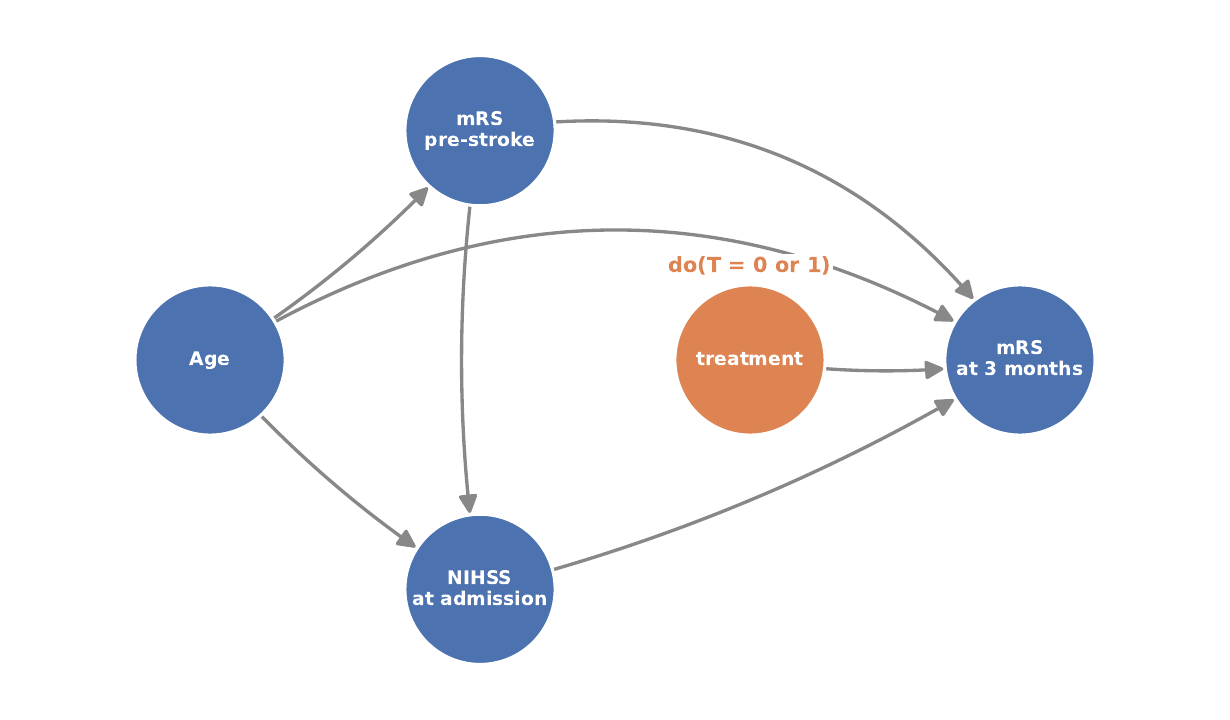}
\caption{Post-interventional DAG after do-intervention on the treatment node:
the incoming edges to treatment are removed while all fitted relationships are kept.}
\label{fig:do-dag}
\end{figure}

Then we used the trained TRAM-DAG model to predict the 
individualized treatment effects (ITE) for all patients in the MR CLEAN RCT cohort (see Figure~\ref{fig:do-dag}). 
To do so for an individual patient of the RCT cohort, 
we fix the values of the causal parent nodes 
of the outcome mRS\_3m beside treatment 
(i.e., age, mRS\_pre, NIHSSa - see Figure~\ref{fig:dag})   to 
the observed values of this patient. To simulate interventions with our TRAM-DAG, we delete all 
arrows pointing to the treatment node and fix the treatment value once with $T=0$ 
and once with $T=1$ (see Figure~\ref{fig:do-dag}). 
We then sample from the TRAM-DAG with the post-interventional DAG structure (see Figure~\ref{fig:do-dag}). This yields an interventional probability distribution of the 
outcome mRS\_3m under $T=0$ 
and $T=1$. Based on these predictions, we can compute the predicted 
probability for a good outcome 
under both treatment options, allowing us to compute the ITE: 

\begin{equation}
\text{ITE}(x) = P(\text{mRS\_3m}\le 2 \mid \mathrm{do}(T{=}1), x)
             - P(\text{mRS\_3m}\le 2 \mid \mathrm{do}(T{=}0), x),
\end{equation}
where $x$ again represents the values of age, mRS\_pre, NIHSSa.

\begin{figure}[h!]
\centering
\includegraphics[width=0.7\linewidth]{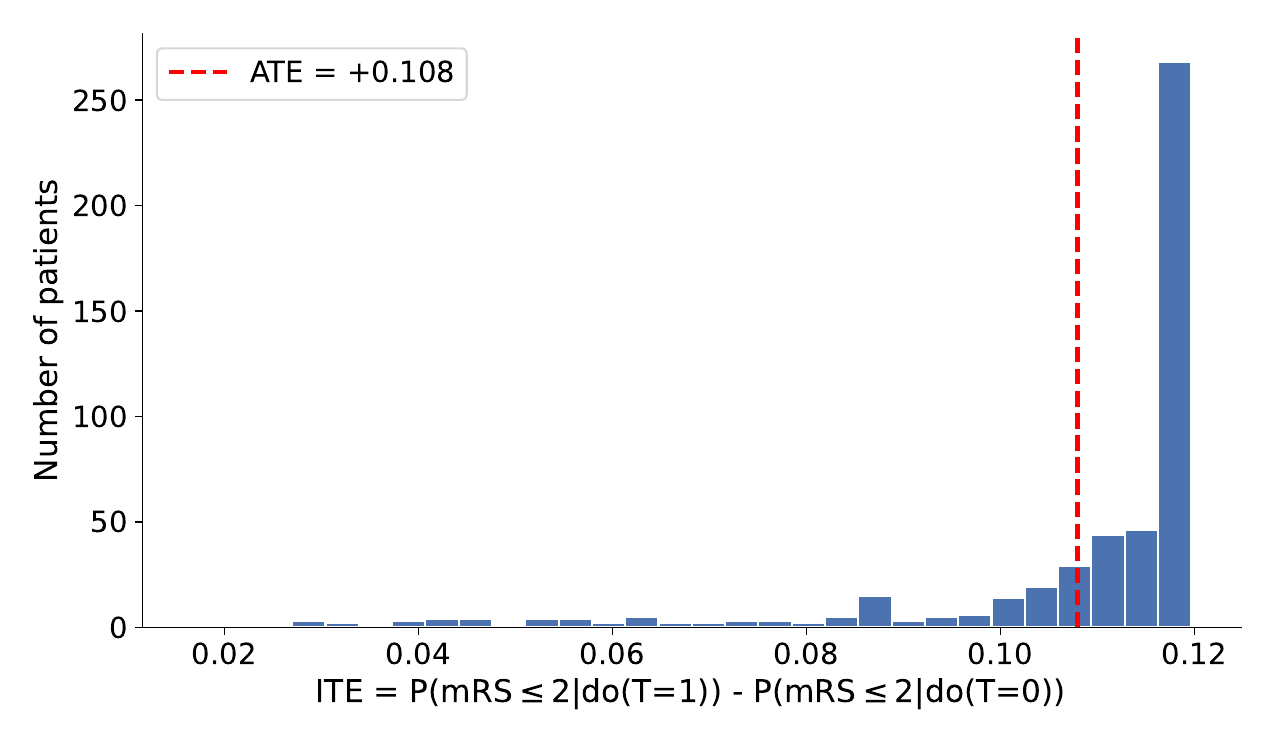}
\caption{Distribution of estimated Individualized Treatment Effects (ITE) on
the probability of a good outcome for the MR CLEAN cohort.}
\label{fig:ite}
\end{figure}

The distribution of the estimated ITEs across patients in the MR CLEAN cohort 
is shown in Figure~\ref{fig:ite}. 
The fact that all ITE predictions are positive (see Figure~\ref{fig:ite}) is 
enforced by modeling a simple homogeneous treatment effect (linear shift effect) as recommended by \cite{belias2021tutorial} for medium sized training data.

\subsection{Validation of the predicted individualized treatment effects}

ITE predictions cannot be directly measured experimentally, 
since each individual patient only receives one treatment. This is known
as the fundamental problem of causal inference.
However, we can check if the predicted ITE of our TRAM-DAG model is consistent
with the RCT results in different ways.

\subsubsection{Consistency with RCT average treatment effect (ATE)}
First, we validate the model's interventional predictions by comparing the
model-implied average treatment effect (ATE) to the ATE observed in the MR CLEAN RCT.
The TRAM-DAG estimate of $ATE=10.8\%$ is computed as the average of the ITE distribution (Figure~\ref{fig:ite}).
The MR CLEAN trial reported good-outcome rates of $19.1\%$ in the control arm
versus $32.6\%$ in the thrombectomy arm, yielding an absolute difference of
$+0.135$ ($95\%$ CI $5.1$ to $21.9$ percentage points), consistent with the TRAM-DAG ATE estimate of $10.8\%$. The reason for the slightly smaller ATE in the TRAM-DAG model prediction compared to the RCT 
might be due to the fact that the model is trained on the observational MAGIC data, 
where the patients age and NIHSSa distribution differ from the MR CLEAN population, even after restricting to NIHSSa $\ge 6$ (see Figure~\ref{fig:eda}). We have trained once on the full MAGIC data, keeping also the less severe strokes and hence a larger mean NIHSSa difference to the RCT data, and received, as expected, an even lower estimate of the ATE. However, this ATE estimate was also consistent with the reported confidence interval from the MR CLEAN study.

\begin{figure}[h!]
\centering
\includegraphics[width=0.7\linewidth]{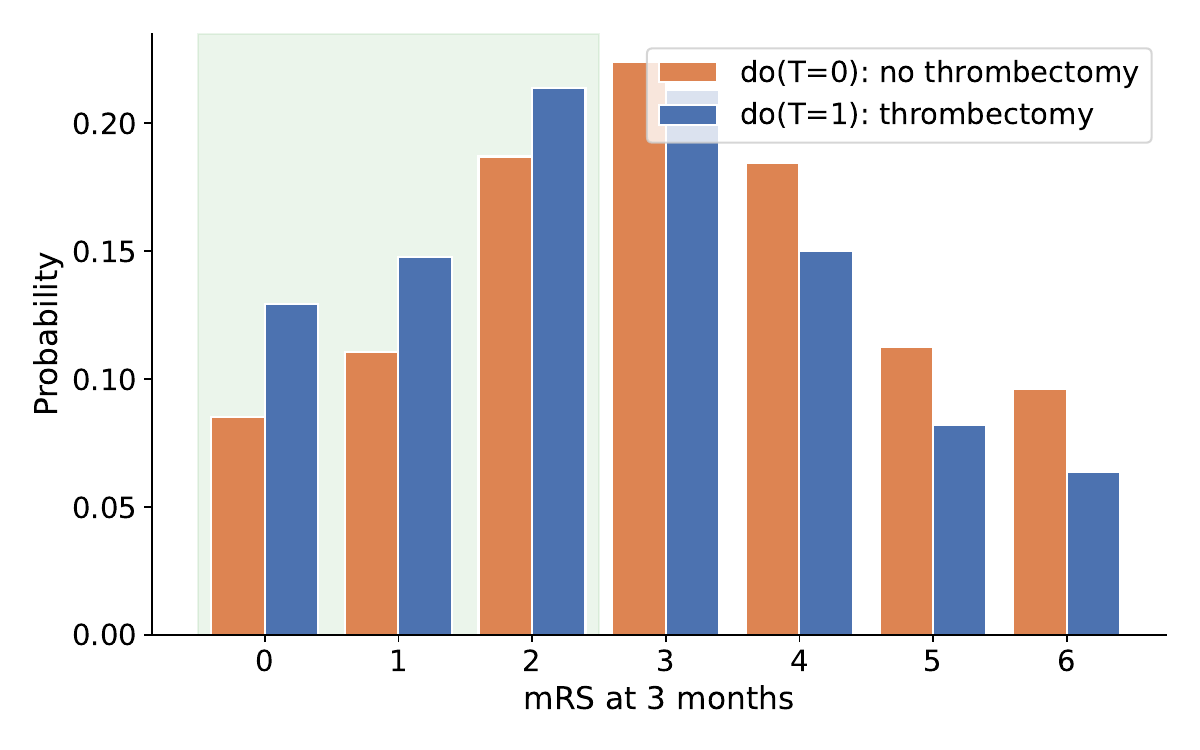}
\caption{Model-simulated RCT: mRS\_3m distribution under $\mathrm{do}(T{=}0)$ and
$\mathrm{do}(T{=}1)$. The shaded region marks a good outcome (mRS\,$\le 2$).
}
\label{fig:rct}
\end{figure}

From the model predictions, we can also compute the predicted distribution 
of mRS\_3m under both treatment arms (see Figure~\ref{fig:rct}), which indicates
a shift towards better outcomes  under treatment, 
with a higher probability of a good outcome (mRS\,$\le 2$) 
in the treated arm compared to the control arm.

\begin{table}[h!]
\centering
\caption{Good outcome probabilities (mRS\,$\le 2$) by treatment arm.
Confidence intervals (CI) for the MR CLEAN observed frequencies are taken
from the original trial publication \citep{mrclean}.
}
\label{tab:ate}
\begin{tabular}{lccc}
\hline
Source & P(good $\mid$ T=0) & P(good $\mid$ T=1) & Difference \\
\hline
TRAM-DAG (simulated) & 0.383 & 0.491 & $+0.108$ \\
MR CLEAN (observed) & 0.191 [0.146, 0.244] & 0.326 [0.267, 0.390] & $+0.135$ [+0.051, +0.219] \\
\hline
\end{tabular}

\end{table}

\subsection{Validation of the predicted probabilities for good outcome}
\begin{figure}[h!]
\centering
\includegraphics[width=0.6\linewidth]{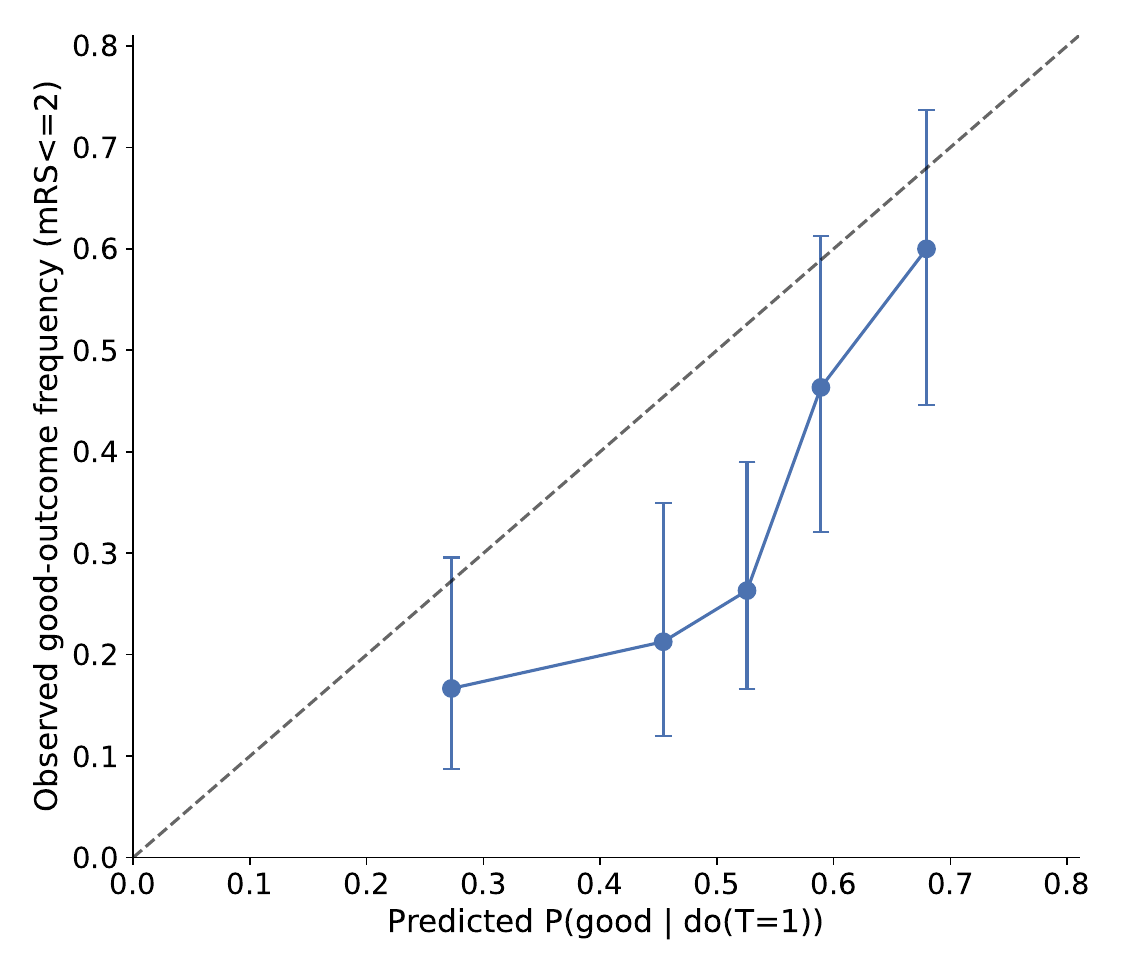}
\caption{Discrimination in the treated arm ($T=1$): predicted vs. observed
good-outcome frequency, binned by predicted probability.}
\label{fig:discr}
\end{figure}

As discussed, we cannot validate the predicted ITE directly; however, we can check the predicted probabilities for a good outcome 
under the treatment that the patient actually received. 
To do so, we use the  TRAM-DAG model to make predictions under $T=1$ and then assess the discrimination power in the treated arm ($T=1$) 
of the RCT.
To check the discrimination power, we divide the treated patients in the RCT arm
into five quantiles according to their predicted probability of a good 
outcome. 
As shown in Figure~\ref{fig:discr}, patients for whom the TRAM-DAG predicted
probabilities of success indeed experienced good outcomes more frequently.  This suggests that the relative
probabilities used to compute the ITE are  trustworthy for identifying
which patients have the best chance of recovery.

However, the model's absolute probabilities for a good outcome are not yet perfectly calibrated but tend to be slightly over-optimistic. This might again be due to the fact that the model is trained on observational data, which have a different distribution than the RCT data(see Figure~\ref{fig:eda}).

\section{Discussion}
\label{sec:disc}
A TRAM-DAG fitted to MAGIC multi-center observational stroke data (filtered for NIHSSa $>=6$) reproduces the
direction and approximate magnitude of the thrombectomy benefit established by
MR CLEAN and ranks treated patients by outcome correctly, even if the predicted probability for a good outcome is slightly over-optimistic.

In future work, we will include more observational and RCT, which should allow us 
to explore more flexible models, including heterogeneous treatment effects and a more rigorous evaluation.  We also aim to investigate how potentially unobserved confounders could be handled.

\appendix

\bibliographystyle{plainnat}
\bibliography{references}

@article{mrclean,
  title   = {A Randomized Trial of Intraarterial Treatment for Acute Ischemic Stroke},
  author  = {Berkhemer, Olvert A. and Fransen, Puck S. S. and Beumer, Debbie and
             van den Berg, Lucie A. and Lingsma, Hester F. and Yoo, Albert J. and
             Schonewille, Wouter J. and Vos, Jan Albert and Nederkoorn, Paul J. and
             Wermer, Marieke J. H. and others},
  journal = {New England Journal of Medicine},
  volume  = {372},
  number  = {1},
  pages   = {11--20},
  year    = {2015},
  doi     = {10.1056/NEJMoa1411587},
  note    = {MR CLEAN Investigators}
}

@article{magic,
  title   = {The Multicentre Acute ischemic stroke imaGIng and Clinical data ({MAGIC}) repository: rationale and blueprint},
  author  = {Baazaoui, H. and Engelter, S. T. and Gensicke, H. and Enz, L. S. and
             Psychogios, M. and Mutke, M. and Michel, P. and Strambo, D. and
             Salerno, A. and Marquering, H. A. and Nederkoorn, P. J. and Wali, N. and
             Tanadini-Lang, S. and Menze, B. and {de la Rosa}, E. and Yang, K. and
             {De Marchis}, G. M. and Dittrich, T. D. and Valletta, F. and Germann, M. and
             Cereda, C. W. and Marto, J. P. and Herzog, L. and Hirschi, P. and
             Kulcsar, Z. and Wegener, S.},
  journal = {Frontiers in Neuroinformatics},
  volume  = {18},
  pages   = {1508161},
  year    = {2025},
  doi     = {10.3389/fninf.2024.1508161}
}

@article{tramdag,
  title   = {Interpretable Neural Causal Models with {TRAM-DAGs}},
  author  = {Sick, Beate and D{\"u}rr, Oliver},
  journal = {Proceedings of Machine Learning Research},
  volume  = {275},
  pages   = {1--25},
  year    = {2025}
}

@article{ontram,
  title   = {Deep and interpretable regression models for ordinal outcomes},
  author  = {Kook, L. and Herzog, L. and Hothorn, T. and D{\"u}rr, O. and Sick, B.},
  journal = {Pattern Recognition},
  volume  = {122},
  pages   = {108263},
  year    = {2022},
  doi     = {10.1016/j.patcog.2021.108263}
}

@article{glip,
  title={Exact Graph Learning via Integer Programming},
  author={Kook, Lucas and Mogensen, S{\o}ren Wengel},
  journal={arXiv preprint arXiv:2601.20589},
  year={2026}
}

@article{belias2021tutorial,
title={A tutorial on individualized treatment effect prediction from randomized trials with a binary endpoint},
  author={Hoogland, Jeroen and IntHout, Joanna and Belias, Michail and Rovers, Maroeska M and Riley, Richard D and E. Harrell Jr, Frank and Moons, Karel GM and Debray, Thomas PA and Reitsma, Johannes B},
  journal={Statistics in medicine},
  volume={40},
  number={26},
  pages={5961--5981},
  year={2021},
  publisher={Wiley Online Library}
}

\end{document}